# A matrix solution to Maxwell’s equations in 2 + 1 dimensional curved space: Two examples

S.G.Kamath* [0000-0002-1148-3787]

Department of Mathematics, Indian Institute of Technology Tirupati

Renigunta Road, Tirupati 517506,India

*e-mail:gkamath01865@gmail.com

**Abstract**: In an effort that puts together a paper by Plebanski[1] with a matrix approach to the solution of Maxwell’s equations in flat space by Moses[2], Maxwell’s equations in 2 + 1 dimensional curved space are solved in two separate cases of the metric $g_{\mu\nu}$ given by:

a. Banados, Teitelboim and Zanelli[3] and b. Deser, Jackiw and ‘tHooft[4] and Clement[5] to obtain the respective time – independent solutions; an extension to time – dependent solutions with the same point of view is also briefly indicated.

The Coulomb potential $\phi = \ln r, r^2 = x^2 + y^2$ is an obvious solution to the Laplace equation $\nabla^2 \phi = 0$, with the general solution easily worked out as

$$\phi(r,\theta) = \sum_{m=0}^{\infty} \left(a_m r^m + b_m r^{-m}\right)\left(c_m \sin m\theta + d_m \cos m\theta\right) \tag{1}$$

While the simplicity of the Coulomb potential is lost in (1), there have been several recent calculations [6,7,8,9]on the correction to the hydrogen spectrum from the inclusion of the Schwarzschild metric as a perturbation for example. In this context a complementary investigation that is wanting is the quantum mechanics of an hydrogen ion subject to an electrostatic potential $\phi(r,\theta)$without the inconvenience of eq. (1) and the matrix solution to the Maxwell's equations in flat space by Moses[2] that yields both the static and time – dependent solutions as an integral given by eqs.(3.9) and (3.14) therein is perhaps an answer. Going further, the restriction to flat space in Moses[2] motivates this report to build on Ref.2 and solve Maxwell's equations in curved space; the answer thus obtained would be the counterpart of the flat space results mentioned above.

In this effort it pays to link the twin ideas of Plebanski[1] and Moses[2] and as a first we solve Maxwell's equations in 2 + 1 dimensional curved space in this paper; the calculation is lengthy and a reader–friendly approach seems in order. With the equations formally given in 3 + 1 dimensions as

$$\begin{gathered}\frac{1}{\sqrt{-g}}\partial_\mu\left(\sqrt{-g}\,F^{\mu\nu}\right) = \frac{4\pi}{c} j^\nu, \nabla_\beta {}^*F^{\alpha\beta} = 0 \\ F^{\mu\nu} = g^{\mu\alpha} g^{\nu\beta} F_{\alpha\beta}, F_{\alpha\beta} = \partial_\beta A_\alpha - \partial_\alpha A_\beta, {}^*F^{\alpha\beta} = \frac{1}{2}\varepsilon^{\alpha\beta\mu\nu} F_{\mu\nu}\end{gathered} \tag{2}$$

it helps to elaborate eq.(2) in 2 + 1 dimensions as the four equations involving partial derivatives namely,

$$\begin{aligned} &D_{i,i} = -\frac{4\pi}{c} j^0, && D_{1,0} + H_{,2} = \frac{4\pi}{c} j^1, \\ &D_{2,0} - H_{,1} = \frac{4\pi}{c} j^2, && -B_{,0} + E_{2,1} - E_{1,2} = 0 \end{aligned} \tag{2a}$$

with[1]

$$D_a = \sqrt{-g} F^{0a}, H = \sqrt{-g}\,F^{21}, E_a = F_{a0}, B = F_{21} \tag{3}$$

The last entry in eq.(2a) is the Bianchi identity, and for the examples we shall consider here $g$ is a constant with $|g| = 1$.

With the pair of column vectors defined as

$$\Psi \equiv \begin{pmatrix} B+iH & E_1+iD_1 & E_2+iD_2 & 0 \end{pmatrix}^T, \Gamma \equiv \begin{pmatrix} 0 & \frac{4\pi i}{c} j^1 & \frac{4\pi i}{c} j^2 & \frac{4\pi i}{c} j^0 \end{pmatrix}^T \tag{4}$$

eqs.(2a) can now be written following Ref.2 as

$$\left(I\partial_0 + \alpha_1\partial_1 + \alpha_2\partial_2\right)\Psi = \Gamma \tag{5}$$

with $I$ being the unit matrix 4 x 4 matrix and the $\alpha_i$ defined by

$$\alpha_1 = \begin{pmatrix} 0 & 0 & -1 & 0 \\ 0 & 0 & 0 & 1 \\ -1 & 0 & 0 & 0 \\ 0 & -1 & 0 & 0 \end{pmatrix}, \alpha_2 = \begin{pmatrix} 0 & 1 & 0 & 0 \\ 1 & 0 & 0 & 0 \\ 0 & 0 & 0 & 1 \\ 0 & 0 & -1 & 0 \end{pmatrix} \tag{6}$$

The operator $\vec{\alpha} \cdot \vec{\partial}$ has eigenvalues $\pm p, \pm ip$ with $p = |\vec{p}|$, with the respective linearly independent set of orthonormal eigenvectors being

$$\begin{aligned} \sqrt{2}p\chi_1 &= e^{i\vec{p}\cdot\vec{x}}\begin{pmatrix} 0 & ip_1 & ip_2 & p \end{pmatrix}^T, \quad \sqrt{2}p\chi_2 = e^{i\vec{p}\cdot\vec{x}}\begin{pmatrix} 0 & p_1 & p_2 & ip \end{pmatrix}^T \\ \sqrt{2}p\chi_3 &= e^{i\vec{p}\cdot\vec{x}}\begin{pmatrix} p & p_2 & -p_1 & 0 \end{pmatrix}^T, \sqrt{2}p\chi_4 = e^{i\vec{p}\cdot\vec{x}}\begin{pmatrix} -p & p_2 & -p_1 & 0 \end{pmatrix}^T \end{aligned} \tag{7}$$

These eigenvectors satisfy the orthonormality and the completeness property respectively given by

$$\int_p \chi_i^+(x)\chi_j(y) = \delta_{ij}\delta^{(2)}(\vec{x}-\vec{y}), \sum_{i=1}^{4}\int_p \chi_{ia}(\vec{x},\vec{p})\chi_{ib}^*(\vec{y},\vec{p}) = \delta_{ab}\delta^{(2)}(\vec{x}-\vec{y}) \tag{8}$$

The subscripts $a,b$ in eq.(8) label the elements of the column vectors $\chi_i$ with the operation of complex(hermitian) conjugation shown as $\chi_{ib}^*$ $\left(\chi_i^+\right)$.

A diligent application of the work by Moses[2] now helps to determine the matrix elements in the column vector

$$\Psi = \begin{pmatrix} B+iH & E_1+iD_1 & E_2+iD_2 & 0 \end{pmatrix}^T \tag{9}$$

in terms of those in

$$\Gamma = \begin{pmatrix} 0 & \frac{4\pi i}{c} j^1 & \frac{4\pi i}{c} j^2 & \frac{4\pi i}{c} j^0 \end{pmatrix}^T \tag{10}$$

from the expansion

$$\Psi = \sum_{i=1}^{4}\int_p \chi_i \, h_i(\vec{p},t) \, , \Gamma = \sum_{i=1}^{4}\int_p \chi_i \; g_i(\vec{p},t), \quad \int_p \equiv \int_{-\infty}^{\infty} \frac{d^2 p}{(2\pi)^2} \tag{11}$$

With

$$\vec{\alpha}\cdot\vec{\partial}\,\Psi = \int_p \left(p\chi_1 h_1 - p\chi_2 h_2 + ip\chi_3 h_3 - ip\chi_4 h_4\right) \tag{12}$$

the **time – independent** solutions to (5) can now be obtained in two steps using eqs.(8),(12) through

$$\vec{\alpha}\cdot\vec{\partial}\Psi = \Gamma \tag{13}$$

One first gets

$$ph_1 = g_1, ph_2 = -g_2, ph_3 = -ig_3, ph_4 = ig_4 \tag{14}$$

and the use of eqs.(14) and then eqs.(8) enables the rewrite

$$\Psi(x) = \int_p\int_z \frac{1}{p}\left(\chi_1\chi_1^+ - \chi_2\chi_2^+ - i\chi_3\chi_3^+ + i\chi_1\chi_1^+\right)\Gamma(\vec{z}) \tag{15}$$

with the $\chi_i$ and their hermitian conjugates being functions of $\vec{x},\vec{z}$ respectively. With eqs.(7), eq.(15) reads as

$$\Psi(x) = \int_p\int_z \frac{e^{i\,\vec{p}\cdot(\vec{x}-\vec{z})}}{2p^3}\begin{pmatrix} 0 & -2ipp_2 & 2ipp_1 & 0 \\ -2ipp_2 & 0 & 0 & 2ipp_1 \\ 2ipp_1 & 0 & 0 & 2ipp_2 \\ 0 & -2ipp_1 & -2ipp_2 & 0 \end{pmatrix}\Gamma(\vec{z}) \tag{16}$$

Or,

$$\Psi(x) = \frac{4\pi i}{c}\int_p\int_z \frac{e^{i\vec{p}\cdot(\vec{x}-\vec{z})}}{2p^3}\left(-2ip\left(p_2 j^1 - p_1 j^2\right) \;\; 2ipp_1 j^0 \;\; 2ipp_2 j^0 \;\; -2ip\left(p_1 j^1 + p_2 j^2\right)\right)^T$$

The last term will be zero by current conservation and the momentum integration then leads to

$$\Psi(x) = \frac{2i}{c}\int_z \left( \frac{\vec{j}\times(\vec{x}-\vec{z})}{|\vec{x}-\vec{z}|^2} \quad -\frac{j^0}{2}\partial_x \log|\vec{x}-\vec{z}|^2 \quad -\frac{j^0}{2}\partial_y \log|\vec{x}-\vec{z}|^2 \quad 0 \right)^T \tag{17}$$

Equating this to

$$\Psi = \begin{pmatrix} B+iH & E_1+iD_1 & E_2+iD_2 & 0 \end{pmatrix}^T$$

it is easy to infer that

$$\vec{E}+i\vec{D} = -i\ grad\left(\log\left|\vec{x}-\vec{R}\right|^2\right) \tag{18}$$

when $j^0 = c\delta\left(\vec{z}-\vec{R}\right)$ reflecting an unit charge at $\vec{z}=\vec{R}$. By extension this implies that

$$B+iH = \frac{2i}{c}\int_z \frac{\vec{j}\times\left(\vec{x}-\vec{z}\right)}{\left|\vec{x}-\vec{z}\right|^2} \tag{18a}$$

The relation between $\left(B,\vec{E}\right)$ and $\left(H,\vec{D}\right)$ now follows from

$$F_{a0} = g_{ab}g_{0c}F^{bc}, F_{21} = g_{2\mu}g_{1\nu}F^{\mu\nu} \tag{19}$$

and eqs.(3).In detail, one has

$$F_{a0} = \left(g_{a0}g_{01}-g_{a1}g_{00}\right)F^{01} + \left(g_{a0}g_{02}-g_{a2}g_{00}\right)F^{02} + \left(g_{a2}g_{01}-g_{a1}g_{20}\right)F^{21}$$

$$F_{21} = \left(g_{20}g_{11}-g_{21}g_{10}\right)F^{01} + \left(g_{20}g_{12}-g_{22}g_{10}\right)F^{02} + \left(g_{22}g_{11}-g_{21}g_{12}\right)F^{21}$$

leading to,

$$\begin{aligned} B+iH &\equiv F_{21}+iF^{21} = \left(g_{20}g_{11}-g_{21}g_{10}\right)F^{01} + \left(g_{20}g_{12}-g_{22}g_{10}\right)F^{02} + \left(g_{22}g_{11}-g_{21}g_{12}+i\right)F^{21} \\ E_1+iD_1 &\equiv F_{10}+iF^{01} = \left(g_{10}g_{01}-g_{11}g_{00}+i\right)F^{01} + \left(g_{10}g_{02}-g_{12}g_{00}\right)F^{02} + \left(g_{12}g_{01}-g_{11}g_{20}\right)F^{21} \\ E_2+iD_2 &\equiv F_{20}+iF^{02} = \left(g_{20}g_{01}-g_{21}g_{00}\right)F^{01} + \left(g_{20}g_{02}-g_{22}g_{00}+i\right)F^{02} + \left(g_{22}g_{01}-g_{21}g_{20}\right)F^{21} \end{aligned} \tag{20}$$

Writing eqs.(20) as

$$\begin{aligned} &\begin{pmatrix} B+iH & E_1+iD_1 & E_2+iD_2 \end{pmatrix}^T \\ &= \begin{pmatrix} \left(g_{20}g_{11}-g_{21}g_{10}\right) & \left(g_{20}g_{12}-g_{22}g_{10}\right) & \left(g_{22}g_{11}-g_{21}g_{12}+i\right) \\ \left(g_{10}g_{01}-g_{11}g_{00}+i\right) & \left(g_{10}g_{02}-g_{12}g_{00}\right) & \left(g_{12}g_{01}-g_{11}g_{20}\right) \\ \left(g_{20}g_{01}-g_{21}g_{00}\right) & \left(g_{20}g_{02}-g_{22}g_{00}+i\right) & \left(g_{22}g_{01}-g_{21}g_{20}\right) \end{pmatrix} \begin{pmatrix} F^{01} \\ F^{02} \\ F^{21} \end{pmatrix} \end{aligned} \tag{21}$$

one can now determine through matrix inversion the column vector on the right hand side of (21) as eqs.(18) and (18a) define the left hand side of eq.(21) for an unit charge at $\vec{z}=\vec{R}$.

This will be done first for the $g_{\mu\nu}$ given in Cartesian coordinates by:

1. The black – hole metric of M.Banados, C.Teitelboim and J.Zanelli[3]:

$$g_{\mu\nu} = \begin{pmatrix} M - \frac{r^2}{l^2} & \frac{Jy}{2r^2} & -\frac{Jx}{2r^2} \\ \frac{Jy}{2r^2} & \frac{x^2 + y^2 f^2}{r^2 f^2} & \frac{xy}{r^2}\left(\frac{1}{f^2} - 1\right) \\ -\frac{Jx}{2r^2} & \frac{xy}{r^2}\left(\frac{1}{f^2} - 1\right) & \frac{y^2 + x^2 f^2}{r^2 f^2} \end{pmatrix} \tag{22a}$$

$$f^2 = -M + \frac{r^2}{l^2} + \frac{J^2}{4r^2}, \quad r^2 = x^2 + y^2$$

eq.(22a) being obtained from

$$ds^2 = -\left(N^{\perp}\right)^2 dt^2 + \frac{1}{f^2} dr^2 + r^2\left(d\theta + N^{\theta} dt\right)^2, 2r^2 N^{\theta} = -J, \left(N^{\perp}\right)^2 = f^2 = -M + \frac{r^2}{l^2} + \frac{J^2}{4r^2} \tag{22b}$$

In this case eq.(21) reads as

$$\left(B + iH \quad E_1 + iD_1 \quad E_2 + iD_2\right)^T$$

$$= L\left(F^{01}\ F^{02}\ F^{21}\right)^T$$

$$= \begin{pmatrix} -\frac{Jx}{2r^2 f^2} & -\frac{Jy}{2r^2 f^2} & \frac{1}{f^2} + i \\ \frac{1}{r^2}\left(y^2 f^2 + \frac{x^2}{f^2}\left(-M + \frac{r^2}{f^2}\right)\right) + i & -\frac{xy}{r^2}\left(f^2 + \frac{1}{f^2}\left(M - \frac{r^2}{f^2}\right)\right) & \frac{Jx}{2r^2 f^2} \\ -\frac{xy}{r^2}\left(f^2 + \frac{1}{f^2}\left(M - \frac{r^2}{f^2}\right)\right) & \frac{1}{r^2}\left(x^2 f^2 + \frac{y^2}{f^2}\left(-M + \frac{r^2}{f^2}\right)\right) + i & \frac{Jy}{2r^2 f^2} \end{pmatrix} \begin{pmatrix} F^{01} \\ F^{02} \\ F^{21} \end{pmatrix} \tag{23}$$

The matrix $L$ is invertible as

$$\det L = \left(1 + \frac{i}{f^2}\right)\left\{1 + i\left(-M + \frac{r^2}{l^2}\right)\right\} - \left(f^2 + \frac{1}{f^2}\right) \neq 0 \tag{24}$$

and

$$
(\det L)L^{-1} = \begin{pmatrix} -acx & d\left(i+\frac{f^2x^2}{r^2}\right)+k\frac{y^2}{f^2r^2} & xy\left(d\frac{f^2}{r^2}-k\frac{1}{f^2r^2}\right) \\ -acy & xy\left(d\frac{f^2}{r^2}-k\frac{1}{f^2r^2}\right) & d\left(i+\frac{f^2y^2}{r^2}\right)+k\frac{x^2}{f^2r^2} \\ ic+g\left(-M+\frac{r^2}{l^2}\right) & acx & acy \end{pmatrix} \tag{25}
$$

with
$$
a \equiv \frac{J}{2r^2 f^2}, c \equiv i + f^2, d \equiv i + \frac{1}{f^2}, g \equiv 1 + \frac{i}{f^2}, k \equiv 1 + i\left(-M + \frac{r^2}{l^2}\right) \tag{26}
$$

Eq.(23) thus yields

$$
\left(F^{01} \;\; F^{02} \;\; F^{21}\right)^T = L^{-1}\left(B + iH \quad E_1 + iD_1 \quad E_2 + iD_2\right)^T
$$

$$
= \frac{1}{\det L}\begin{pmatrix} -acx & d\left(i+\frac{f^2x^2}{r^2}\right)+k\frac{y^2}{f^2r^2} & xy\left(d\frac{f^2}{r^2}-k\frac{1}{f^2r^2}\right) \\ -acy & xy\left(d\frac{f^2}{r^2}-k\frac{1}{f^2r^2}\right) & d\left(i+\frac{f^2y^2}{r^2}\right)+k\frac{x^2}{f^2r^2} \\ ic+g\left(-M+\frac{r^2}{l^2}\right) & acx & acy \end{pmatrix}\begin{pmatrix} B+iH \\ E_1+iD_1 \\ E_2+iD_2 \end{pmatrix} \tag{27}
$$

With eqs.(18) and (18a) defining the column vector on the right hand side of (27) it is easy to determine the required answer for the metric given by (22a).

A similar effort informs the second example below:

2. The metric of S.Deser, R.Jackiw and G. 'tHooft[4] and G.Clement[5] is obtained from $ds^2 = (cdt + \lambda d\theta)^2 - dr^2 - r^2 d\theta^2, \lambda = \frac{kJ}{2\pi}, k = 8\pi G, J = \left|\vec{J}\right|$ as:

$$
g_{\mu\nu} = \begin{pmatrix} 1 & -\frac{\lambda y}{x^2+y^2} & \frac{\lambda x}{x^2+y^2} \\ -\frac{\lambda y}{x^2+y^2} & -1+\left(\frac{\lambda y}{x^2+y^2}\right)^2 & -\frac{\lambda^2 xy}{\left(x^2+y^2\right)^2} \\ \frac{\lambda x}{x^2+y^2} & -\frac{\lambda^2 xy}{\left(x^2+y^2\right)^2} & -1+\left(\frac{\lambda x}{x^2+y^2}\right)^2 \end{pmatrix} \tag{28}
$$

$G$ being the gravitational constant and $J$ the spin of the massless particle, eq.(28) being labeled as the rotating solution of Ref.4 by Clement[5]. In this case one first gets

$$\begin{pmatrix} B+iH \\ E_1+iD_1 \\ E_2+iD_2 \end{pmatrix} \equiv L \begin{pmatrix} F^{01} \\ F^{02} \\ F^{21} \end{pmatrix} = \begin{pmatrix} -\frac{\lambda x}{x^2+y^2} & -\frac{\lambda y}{x^2+y^2} & 1+i-\frac{\lambda^2}{x^2+y^2} \\ 1+i & 0 & \frac{\lambda x}{x^2+y^2} \\ 0 & 1+i & \frac{\lambda y}{x^2+y^2} \end{pmatrix} \begin{pmatrix} F^{01} \\ F^{02} \\ F^{21} \end{pmatrix} \tag{29}$$

with

$$\det L = -2(1-i)-(1+i)\frac{\lambda^2}{x^2+y^2}\left(2\frac{y^2}{r^2}+i\right) \tag{30}$$

and

$$\begin{pmatrix} F^{01} \\ F^{02} \\ F^{21} \end{pmatrix} = \frac{1}{\det L} \begin{pmatrix} -(1+i)A & 2i+\left(-i-\frac{x^2}{x^2+y^2}\right)C & -AG \\ -(1+i)G & -AG & 2i+\left(-i-\frac{y^2}{x^2+y^2}\right)C \\ 2i & (1+i)A & (1+i)G \end{pmatrix} \begin{pmatrix} B+iH \\ E_1+iD_1 \\ E_2+iD_2 \end{pmatrix} \tag{31}$$

where $$A \equiv \frac{\lambda x}{x^2+y^2}, G \equiv \frac{\lambda y}{x^2+y^2}, C \equiv \frac{\lambda^2}{x^2+y^2} \tag{32}$$

Eqs.(31) and (32) together are the counterpart of (27) for this metric and this completes the derivation of the **time – independent** solutions to eqs.(2).

From eq.(5) the **time – dependent** solutions can also be determined as above from

$$\begin{aligned} &(I\partial_0+\alpha_1\partial_1+\alpha_2\partial_2)\Psi \\ &= \int_p \{\chi_1(ph_1+\partial_0 h_1)+\chi_2(-ph_2+\partial_0 h_2)+\chi_3(iph_3+\partial_0 h_3)+\chi_4(-iph_4+\partial_0 h_4)\} = \sum_{i=1}^{4}\int_p \chi_i g_i \end{aligned} \tag{33}$$

and the four counterparts of eqs.(14)

$$\frac{\partial h_1}{\partial t}+ph_1=g_1,\frac{\partial h_2}{\partial t}-ph_2=g_2,\frac{\partial h_3}{\partial t}+iph_3=g_3,\frac{\partial h_4}{\partial t}-iph_4=g_4 \tag{34}$$

Eqs.(34) have the respective solutions

$$\begin{aligned} h_1(\vec{p},t)&=\int dse^{-p(t-s)}g_1(\vec{p},s), h_2(\vec{p},t)=\int dse^{-p(s-t)}g_2(\vec{p},s)\\ h_3(\vec{p},t)&=\int dse^{-ip(t-s)}g_3(\vec{p},s), h_4(\vec{p},t)=\int dse^{-ip(s-t)}g_4(\vec{p},s)\end{aligned} \tag{35}$$

Eq.(15) now writes as

$$\begin{aligned}\Psi(\vec{x},t)&=\sum_i\int_p \chi_i h_i\\ &=\int_p\int_z\left\{\int ds\, e^{-p(t-s)}\chi_1\chi_1^+ +\int ds\, e^{-p(s-t)}\chi_2\chi_2^+ +\int dse^{-ip(t-s)}\chi_3\chi_3^+ +\int dse^{-ip(s-t)}\chi_4\chi_4^+\right\}\Gamma(\vec{z})\end{aligned} \tag{36}$$

with $\chi_i,\chi_i^+$ being functions of $\vec{x},\vec{z}$ respectively.

For the integration over $p,s$, the first term in (36) yields:

$$\int_p\int ds\frac{e^{i\vec{p}\cdot(\vec{x}-\vec{z})}}{2p^2}\begin{pmatrix}0&0&0&0\\0&p_1^2&p_1p_2&ip_1p\\0&p_1p_2&p_2^2&ip_2p\\0&-ip_1p&-ip_2p&p^2\end{pmatrix}e^{-p(t-s)}\Gamma(\vec{z})\equiv\int ds\int_0^\infty\frac{pdp}{4\pi}e^{-p(t-s)}K\,\Gamma(\vec{z})$$

$$K=\begin{pmatrix}0&0&0&0\\0&\frac{1}{2}(J_0(a)-J_2(a)\cos2\phi)&-\frac{1}{2}J_2(a)\sin2\phi&-J_1(a)\cos\phi\\0&-\frac{1}{2}J_2(a)\sin2\phi&\frac{1}{2}(J_0(a)+J_2(a)\cos2\phi)&-J_1(a)\sin\phi\\0&J_1(a)\cos\phi&J_1(a)\sin\phi&J_0(a)\end{pmatrix} \tag{37}$$

with $a=p|\vec{x}-\vec{z}|,\tan\phi=\frac{x_1-z_1}{x_2-z_2}$ in (37).

From the integrals:

$$\begin{aligned}&\int_0^\infty dx\, xe^{-bx}J_0(ax)=\frac{b}{(b^2+a^2)^{\frac{3}{2}}},b>0\quad;\int_0^\infty dx\, xe^{-bx}J_1(ax)=\frac{a}{(b^2+a^2)^{\frac{3}{2}}},b>0\\ &\int_0^\infty dx\, xe^{-bx}J_2(ax)=\frac{2}{a^2}-\frac{a(2b^2+3a^2)}{a^2(b^2+a^2)^{\frac{3}{2}}},b>0\end{aligned} \tag{38}$$

and the labels

$$\Delta \equiv \left[t^2 + |\vec{x}-\vec{z}|^2\right]^{\frac{1}{2}}, |\vec{x}-\vec{z}|^2 H \equiv (\Delta - t), G \equiv \frac{t}{|\vec{x}-\vec{z}|}, R = \frac{1}{|\vec{x}-\vec{z}|} - \frac{1}{\Delta}$$
$$K \equiv \frac{\sin^2\phi}{|\vec{x}-\vec{z}|} - \frac{\cos^2\phi}{\Delta}, L \equiv \frac{\cos^2\phi}{|\vec{x}-\vec{z}|} - \frac{\sin^2\phi}{\Delta}, 2J \equiv \frac{1}{|\vec{x}-\vec{z}|} + \frac{1}{\Delta} \tag{39}$$

one gets after the integration in (37) the result

$$\frac{1}{4\pi}\begin{pmatrix} 0 & 0 & 0 & 0 \\ 0 & (K + H\cos 2\phi) & (H - J)\sin 2\phi & \frac{\partial}{\partial x}\sinh^{-1} G \\ 0 & (H - J)\sin 2\phi & (L - H\cos 2\phi) & \frac{\partial}{\partial y}\sinh^{-1} G \\ 0 & -\frac{\partial}{\partial x}\sinh^{-1} G & -\frac{\partial}{\partial y}\sinh^{-1} G & R \end{pmatrix} \tag{40}$$

For the remaining terms in (36) one obtains successively:

2nd term:

$$\int_p\int ds \frac{e^{i\vec{p}\cdot(\vec{x}-\vec{z})}}{2p^2}\begin{pmatrix} 0 & 0 & 0 & 0 \\ 0 & p_1^2 & p_1 p_2 & -ip_1 p \\ 0 & p_1 p_2 & p_2^2 & -ip_2 p \\ 0 & ip_1 p & ip_2 p & p^2 \end{pmatrix} e^{-p(s-t)}\Gamma(\vec{z}) \equiv \int ds \int_0^\infty \frac{pdp}{4\pi} e^{-p(s-t)} L\,\Gamma(\vec{z})$$
$$L = \begin{pmatrix} 0 & 0 & 0 & 0 \\ 0 & \frac{1}{2}(J_0(a) - J_2(a)\cos 2\phi) & -\frac{1}{2}J_2(a)\sin 2\phi & J_1(a)\cos\phi \\ 0 & -\frac{1}{2}J_2(a)\sin 2\phi & \frac{1}{2}(J_0(a) + J_2(a)\cos 2\phi) & J_1(a)\sin\phi \\ 0 & -J_1(a)\cos\phi & -J_1(a)\sin\phi & J_0(a) \end{pmatrix} \tag{41}$$

The integration in eq.(41) yields

$$\frac{1}{4\pi}\begin{pmatrix} 0 & 0 & 0 & 0 \\ 0 & (K+E\cos 2\phi) & (E-J)\sin 2\phi & \frac{\partial}{\partial x}\sinh^{-1}G \\ 0 & (E-J)\sin 2\phi & (L-E\cos 2\phi) & \frac{\partial}{\partial y}\sinh^{-1}G \\ 0 & -\frac{\partial}{\partial x}\sinh^{-1}G & -\frac{\partial}{\partial y}\sinh^{-1}G & R \end{pmatrix} \tag{42}$$

with $|\vec{x}-\vec{z}|^2 E = \Delta + t$ as opposed to $|\vec{x}-\vec{z}|^2 H \equiv (\Delta - t)$ in eq.(42).Thus the sum of eqs.(40) and (42) works to

$$\frac{1}{2\pi}\begin{pmatrix} 0 & 0 & 0 & 0 \\ 0 & (K+T\cos 2\phi) & (T-J)\sin 2\phi & \frac{\partial}{\partial x}\sinh^{-1}G \\ 0 & (T-J)\sin 2\phi & (L-T\cos 2\phi) & \frac{\partial}{\partial y}\sinh^{-1}G \\ 0 & -\frac{\partial}{\partial x}\sinh^{-1}G & -\frac{\partial}{\partial y}\sinh^{-1}G & R \end{pmatrix} \tag{43}$$

with $|\vec{x}-\vec{z}|^2 T \equiv \Delta$ .Continuing, one gets for the

3rd term:

$$\int_p\int_t^{\infty} ds \frac{e^{i\vec{p}\cdot(\vec{x}-\vec{z})}}{2p^2}\begin{pmatrix} p^2 & pp_2 & -pp_1 & 0 \\ pp_2 & p_2^2 & -p_1p_2 & 0 \\ -pp_1 & -p_1p_2 & p_1^2 & 0 \\ 0 & 0 & 0 & 0 \end{pmatrix} e^{-ip(s-t)}\Gamma(\vec{z}) \equiv \int_t^{\infty} ds \int_0^{\infty} \frac{pdp}{4\pi} e^{-ip(s-t)} J\,\Gamma(\vec{z}) \tag{44}$$

$$J = \begin{pmatrix} J_0(a) & iJ_1(a)\sin\phi & -iJ_1(a)\cos\phi & 0 \\ iJ_1(a)\sin\phi & \frac{1}{2}(J_0(a)+J_2(a)\cos 2\phi) & \frac{1}{2}J_2(a)\sin 2\phi & 0 \\ -iJ_1(a)\cos\phi & \frac{1}{2}J_2(a)\sin 2\phi & \frac{1}{2}(J_0(a)-J_2(a)\cos 2\phi) & 0 \\ 0 & 0 & 0 & 0 \end{pmatrix}$$

and the integration in Eq.(43) leads to

$$\frac{1}{4\pi}\begin{pmatrix} -iW & -i\frac{\partial}{\partial y}\sin^{-1}\frac{t}{|\vec{x}-\vec{z}|} & i\frac{\partial}{\partial x}\sin^{-1}\frac{t}{|\vec{x}-\vec{z}|} & 0 \\ -i\frac{\partial}{\partial y}\sin^{-1}\frac{t}{|\vec{x}-\vec{z}|} & -iL+S\cos 2\phi & (S-iJ)\sin 2\phi & 0 \\ i\frac{\partial}{\partial x}\sin^{-1}\frac{t}{|\vec{x}-\vec{z}|} & (S-iJ)\sin 2\phi & -iK-S\cos 2\phi & 0 \\ 0 & 0 & 0 & 0 \end{pmatrix} \tag{45}$$

with $$\tilde{\Delta} \equiv \left(|\vec{x}-\vec{z}|^2 - t^2\right)^{\frac{1}{2}}, |\vec{x}-\vec{z}|^2 S \equiv t + i\tilde{\Delta}, W = \frac{1}{|\vec{x}-\vec{z}|} - \frac{1}{\tilde{\Delta}} \tag{46}$$

4th term: Using $|\vec{x}-\vec{z}|^2 T \equiv t - i\tilde{\Delta}$ below one obtains

$$\int_p\int_t^{\infty} ds \frac{e^{i\vec{p}\cdot(\vec{x}-\vec{z})}}{2p^2}\begin{pmatrix} p^2 & -pp_2 & pp_1 & 0 \\ -pp_2 & p_2^2 & -p_1p_2 & 0 \\ pp_1 & -p_1p_2 & p_1^2 & 0 \\ 0 & 0 & 0 & 0 \end{pmatrix} e^{-ip(s-t)}\Gamma(\vec{z}) \equiv \int_t^{\infty} ds \int_0^{\infty}\frac{pdp}{4\pi}e^{-ip(s-t)}L\,\Gamma(\vec{z})$$

$$L = \begin{pmatrix} J_0(a) & -iJ_1(a)\sin\phi & iJ_1(a)\cos\phi & 0 \\ -iJ_1(a)\sin\phi & \frac{1}{2}\left(J_0(a)+J_2(a)\cos 2\phi\right) & \frac{1}{2}J_2(a)\sin 2\phi & 0 \\ iJ_1(a)\cos\phi & \frac{1}{2}J_2(a)\sin 2\phi & \frac{1}{2}\left(J_0(a)-J_2(a)\cos 2\phi\right) & 0 \\ 0 & 0 & 0 & 0 \end{pmatrix} \tag{47}$$

On completing the integration eq.(47) yields

$$\frac{1}{4\pi}\begin{pmatrix} iW & i\frac{\partial}{\partial y}\sin^{-1}\frac{t}{|\vec{x}-\vec{z}|} & -i\frac{\partial}{\partial x}\sin^{-1}\frac{t}{|\vec{x}-\vec{z}|} & 0 \\ i\frac{\partial}{\partial y}\sin^{-1}\frac{t}{|\vec{x}-\vec{z}|} & iL+T\cos 2\phi & (T+iJ)\sin 2\phi & 0 \\ -i\frac{\partial}{\partial x}\sin^{-1}\frac{t}{|\vec{x}-\vec{z}|} & (T+iJ)\sin 2\phi & -iK-T\cos 2\phi & 0 \\ 0 & 0 & 0 & 0 \end{pmatrix} \tag{48}$$

and the sum of eqs.(45) and (48) works to

$$\frac{1}{2\pi}\begin{pmatrix} 0 & 0 & 0 & 0 \\ 0 & \frac{t\cos 2\phi}{|\vec{x}-\vec{z}|^2} & \frac{t\sin 2\phi}{|\vec{x}-\vec{z}|^2} & 0 \\ 0 & \frac{t\sin 2\phi}{|\vec{x}-\vec{z}|^2} & -\frac{t\cos 2\phi}{|\vec{x}-\vec{z}|^2} & 0 \\ 0 & 0 & 0 & 0 \end{pmatrix} \tag{49}$$

Finally, the sum of eqs.(43) and (49) yields

$$\frac{1}{2\pi}\begin{pmatrix} 0 & 0 & 0 & 0 \\ 0 & (K+E\cos 2\phi) & (E-J)\sin 2\phi & \frac{\partial}{\partial x}\sinh^{-1}G \\ 0 & (E-J)\sin 2\phi & (L-E\cos 2\phi) & \frac{\partial}{\partial y}\sinh^{-1}G \\ 0 & -\frac{\partial}{\partial x}\sinh^{-1}G & -\frac{\partial}{\partial y}\sinh^{-1}G & R \end{pmatrix} \tag{50}$$

with $|\vec{x}-\vec{z}|^2 E \equiv (\Delta + t)$.Eq.(36) thus becomes

$$\Psi(\vec{x},t) = \int_z \frac{1}{2\pi}\begin{pmatrix} 0 & 0 & 0 & 0 \\ 0 & (K+E\cos 2\phi) & (E-J)\sin 2\phi & \frac{\partial}{\partial x}\sinh^{-1}G \\ 0 & (E-J)\sin 2\phi & (L-E\cos 2\phi) & \frac{\partial}{\partial y}\sinh^{-1}G \\ 0 & -\frac{\partial}{\partial x}\sinh^{-1}G & -\frac{\partial}{\partial y}\sinh^{-1}G & R \end{pmatrix}\Gamma(\vec{z}) \tag{51}$$

with
$$\begin{aligned} &\Delta \equiv \left[t^2 + |\vec{x}-\vec{z}|^2\right]^{\frac{1}{2}}, G \equiv \frac{t}{|\vec{x}-\vec{z}|}, R = \frac{1}{|\vec{x}-\vec{z}|} - \frac{1}{\Delta}, K \equiv \frac{\sin^2\phi}{|\vec{x}-\vec{z}|} - \frac{\cos^2\phi}{\Delta}, \\ &|\vec{x}-\vec{z}|^2 E = \Delta + t, 2J \equiv \frac{1}{|\vec{x}-\vec{z}|} + \frac{1}{\Delta}, L \equiv \frac{\cos^2\phi}{|\vec{x}-\vec{z}|} - \frac{\sin^2\phi}{\Delta} \end{aligned} \tag{52}$$

and is the counterpart of eq.(14) for the static case.

From eq.(51) it is clear that $B+iH$ will be zero unlike $\vec{E}+i\vec{D}$ and this is therefore one departure from eqs.(17) and (17a); eq.(21) also reminds us that the column vector $\left(F^{01}\ F^{02}\ F^{21}\right)^T$ is got from the column vector $\left(B+iH\ E_1+iD_1\ E_2+iD_2\right)^T$ by inversion and as eqs.(23) and (30) show, one should expect despite $B+iH$ being zero, non – trivial answers for the electric and magnetic fields given by $\left(F^{01}\ F^{02}\ F^{21}\right)^T$; this feature merits a separate discussion and will be taken up elsewhere.

In conclusion, one has determined the solutions to Maxwell's equations in curved space in a form that does not have the infirmity of eq.(1) and can be used to meet the objectives stated in the introduction to this paper; admittedly the presentation has been belaboured given the steps involved.